Effect of the length of inflation on angular TT and TE power spectra in power-law inflation

Shiro Hirai* and Tomoyuki Takami**

Department of Digital Games, Faculty of Information Science and Arts,
Osaka Electro-Communication University
1130-70 Kiyotaki, Shijonawate, Osaka 575-0063, Japan
*Email: hirai@isc.osakac.ac.jp
**Email: takami@isc.osakac.ac.jp

**Abstract**

The effect of the length of inflation on the power spectra of scalar and tensor perturbations is estimated using the power-law inflation model with a scale factor of $a(\eta) = (-\eta)^p = t^q$. Considering various pre-inflation models with radiation-dominated or scalar matter-dominated periods before inflation in combination with two matching conditions, the temperature angular power spectrum (TT) and temperature-polarization cross-power spectrum (TE) are calculated and a likelihood analysis is performed. It is shown that the discrepancies between the Wilkinson Microwave Anisotropy Probe (WMAP) data and the $\Lambda$CDM model, such as suppression of the spectrum at $l = 2,3$ and oscillatory behavior, may be explained by the finite length of inflation model if the length of inflation is near 60 $e$-folds and $q \geq 300$. The proposed models retain similar values of $\chi^2$ to that achieved by the $\Lambda$CDM model with respect to fit to the WMAP data, but display different characteristics of the angular TE power spectra at $l \leq 20$.

**1. Introduction**

Inflation is an important concept in cosmology and is strongly supported by recent satellite-based experiments. However, consistent and natural models of inflation from the point of view of particle physics have yet to be established. Recently released data obtained by the Wilkinson Microwave Anisotropy Probe (WMAP) has allowed a number of cosmological parameters to be fixed precisely. Although the WMAP data can be almost satisfactorily explained by the $\Lambda$CDM model [1], there remains some inconsistency in the suppression of the spectrum at large angular scales ($l = 2,3$), as well as the running of the spectral index and oscillatory behavior in the spectrum. The effect of this small discrepancy on the inflation models and pre-inflation physics is thus an interesting problem.

Many attempts have been made to explain the gap between the WMAP data and the $\Lambda$CDM model. As simple slow-roll inflation models appear to be unable to adequately explain the three features mentioned above [2], double inflation models and more complicated models have been considered [3]. The suppression of the spectrum seen in the WMAP data at large angular scales has been addressed by many models, including a finite-sized universe model with nontrivial topology or a closed universe [4]. Models involving new physics [5-7], a cutoff [8-9], or an initially kinetics-dominated region or some other region [10] have also been presented. In these studies [5-10], special power spectra are derived or assumed and temperature angular power spectra (TT) having suppression at small $l$ are obtained considering the contribution of trans-Planckian physics [11] or the special behavior of inflation at 60 $e$-folds, or explicit [12] or implicit finite inflation described by a pre-inflation model. However, the temperature-polarization cross-power spectrum (TE) has only been obtained in relatively few studies [6,7,9]. The present study, forming part of a series of work [13-15] on the effect of the length of inflation and pre-inflation physics on the density perturbations, examines finite inflation with a pre-inflation state as a potential explanation of the WMAP data using a model developed in a previous paper [13]. In finite inflation, the length of inflation becomes critical, particularly near 60 *e*-folds, affecting suppression of the angular TT spectrum at $l = 2,3$ and oscillatory behavior. The effect of such a model on the angular TT and TE power spectra is discussed, and a likelihood analysis is performed to evaluate the fit to the WMAP data.

The effect of the initial condition in inflation on the power spectrum of curvature perturbations was considered in a previous paper [13]. Based on the physical conditions before inflation, the possibility exists that the initial state of scalar perturbations in inflation is not simply the Bunch-Davies state, but a more general squeezed state [16]. A formula for the power spectrum of curvature perturbations having any initial condition in inflation was derived for this model as a familiar formulation multiplied by a factor indicating the contribution of the initial condition. Subsequent papers [14,15] considered finite inflation models in which inflation began at a certain time, preceded by pre-inflation as a radiation-dominated or scalar matter-dominated period. Calculations of the power spectrum were made for two matching conditions; one in which the gauge potential and its first η-derivative are continuous at the transition point, and one in which the transition occurs on a hyper-surface of constant energy as proposed by Deruelle and Mukhanov [17]. The differences among the models and matching conditions were investigated in detail by calculating the power spectrum of curvature perturbations, the spectral index, and the running spectral index. This analysis revealed that when the length of inflation is finite, the power spectra can be expressed by decreasing functions at super large scales, and that in cases of radiation-dominated and scalar matter-dominated pre-inflation with the special matching condition, the power spectra oscillate from large to small scales.

In this paper, the effect of such properties of the power spectrum on the angular TT and TE power spectra is investigated for these pre-inflation models. The model assumes inflation of finite length in the case of power-law inflation using a scale factor of $a(\eta) = (-\eta)^p = t^q$ with either radiation-dominated or scalar matter-dominated pre-inflation and two matching conditions. These toy pre-inflation models are used for the following reasons. If the universe is very hot in pre-inflation, even if a massive particle dominates the energy density of the universe, the universe can be approximated as being dominated by radiation-like matter. Alternatively, it is natural that the scalar field that causes inflation dominates the energy density of the universe. The power spectra of curvature perturbations and gravitational waves can be calculated analytically for the two models and matching conditions. Using the CMBFAST code [18] and the derived power spectra, the angular TT and TE power spectra are calculated for all cases over various lengths of inflation, values of $p$, and reionization optical depth $\tau$. A likelihood analysis with respect to the fit between the TT and TE spectra and the WMAP data is performed using the WMAP code [19]. Comparison of the derived angular power spectra with the WMAP data and the angular power spectrum of the $\Lambda$CDM model yields some interesting results.

This paper is organized as follows. In section 2, formulae for the power spectra of curvature perturbations and gravitational waves are re-derived for any initial condition in inflation. In section 3, the power spectra for curvature perturbations and gravitational waves are re-derived for the two models. In section 4, using the derived formula, the angular TT and TE power spectra and the values of $\chi^2$ are calculated and compared with WMAP data. In section 5, the results obtained in the present study are discussed at length.

## 2. Scalar and tenser perturbations

The formula for the power spectrum of curvature perturbations in inflation is derived here for any initial condition by applying a commonly used method [20]. This formula was originally derived in Refs. [13,21], but as it represents a critical result, it is derived again here for completeness. As a background spectrum, we consider a spatially flat Friedman-Robertson-Walker (FRW) universe described by metric perturbations. The line element for the background and perturbations is generally expressed as [22]

$$ds^2 = a^2(\eta)\{(1+2A)d\eta^2 - 2\partial_i B dx^i d\eta - [(1-2\Psi)\delta_{ij} + 2\partial_i\partial_j E + h_{ij}]dx^i dx^j\}, \quad (1)$$

where $\eta$ is the conformal time. The functions $A$, $B$, $\Psi$ and $E$ represent the scalar perturbations, and $h_{ij}$ represent tensor perturbations. The density perturbation in terms of the intrinsic curvature perturbation of comoving hypersurfaces is given by $\mathfrak{R} = -\Psi - (H/\dot{\phi})\delta\phi$, where $\phi$ is the inflaton field, $\delta\phi$ is the fluctuation of the inflaton field, $H$ is the Hubble expansion parameter, and $\mathfrak{R}$ is the curvature perturbation. Overdots represent derivatives with respect to time $t$, and the prime represents the derivative with respect to the conformal time η. Introducing the gauge-invariant potential $u \equiv a(\eta)(\delta\phi + (\dot{\phi}/H)\Psi)$ allows the action for scalar perturbations to be written as [23]

$$S = \frac{1}{2}\int d\eta d^3x \left\{ \left(\frac{\partial u}{\partial \eta}\right)^2 - c_s^2 (\nabla u)^2 + \frac{Z''}{Z} u^2 \right\}, \qquad (2)$$

where $c_s$ is the velocity of sound, $Z = a\dot{\phi}/H$, and $u = -Z\mathfrak{R}$. Next, we consider the tensor perturbations, where $h_{ij}$ represents the gravitational waves. Under the transverse traceless gauge, the action of gravitational waves in the linear approximation is given by [22]

$$S = \frac{1}{2}\int d^4x \left\{ \left(\frac{\partial h}{\partial \eta}\right)^2 - (\nabla h)^2 + \frac{a''}{a} h^2 \right\}, \qquad (3)$$

where $h$ is the transverse traceless part of the deviation of $h_{ij}$ and represents the two independent polarization states of the wave ($h_+, h_\times$). The fields $u(\eta,x)$ and $h(\eta,x)$ are expressed using annihilation and creation operators as follows.

$$u(\eta,x) = \frac{1}{(2\pi)^{3/2}}\int d^3k \left\{ u_k(\eta) \boldsymbol{a_k} + u_k^*(\eta) \boldsymbol{a_{-k}}^\dagger \right\} e^{-ikx}, \qquad (4)$$

$$h(\eta,x) = \frac{1}{(2\pi)^{3/2} a(\eta)}\int d^3k \left\{ v_k(\eta) \boldsymbol{a_k} + v_k^*(\eta) \boldsymbol{a_{-k}}^\dagger \right\} e^{-ikx}. \qquad (5)$$

The field equation for $u_k(\eta)$ is derived as

$$\frac{d^2 u_k}{d\eta^2} + \left( c_s^2 k^2 - \frac{1}{Z}\frac{d^2 Z}{d\eta^2} \right) u_k = 0. \qquad (6)$$

The field equation for $v_k(\eta)$ becomes equation (6) with $c_s^2 = 1$ and $Z = a(\eta)$. The solution to $u_k$ and $v_k$ satisfy the normalization condition $u_k du_k^*/d\eta - u_k^* du_k/d\eta = i$. For power-law inflation described by $a(\eta) \approx (-\eta)^p \,(= t^{p/(p+1)})$, equation (6) is rewritten as

$$\frac{d^2 u_k}{d\eta^2} + \left(k^2 - \frac{p(p-1)}{\eta^2}\right) u_k = 0, \tag{7}$$

where $c_s^2 = 1$ in the scalar field case. The solution for equation (7) is then written as

$$f_k^I(\eta) = i\frac{\sqrt{\pi}}{2} e^{-ip\pi/2} (-\eta)^{1/2} H^{(1)}_{-p+1/2}(-k\eta), \tag{8}$$

where $H^{(1)}_{-p+1/2}$ is the Hankel function of the first kind with order $-p + 1/2$. As a general initial condition, the mode functions $u_k(\eta)$ and $v_k(\eta)$ are assumed to be

$$u_k(\eta) = c_1 f_k^I(\eta) + c_2 f_k^{I*}(\eta), \tag{9}$$

$$v_k(\eta) = c_{g1} f_k^I(\eta) + c_{g2} f_k^{I*}(\eta), \tag{10}$$

where the coefficients $c_1$ and $c_2$ ($c_{g1}$ and $c_{g2}$) obey the relation $|c_1|^2 - |c_2|^2 = 1$ ($|c_{g1}|^2 - |c_{g2}|^2 = 1$). The important point here is that the coefficients $c_1$ and $c_2$ ($c_{g1}$ and $c_{g2}$) do not change during inflation. In ordinary cases, the field $u_k(\eta)$ is considered to be in the Bunch-Davies state, i.e., $c_1 = 1$ and $c_2 = 0$, because as $\eta \to -\infty$, the field $u_k(\eta)$ must approach plane waves, e.g., $e^{-ik\eta}/\sqrt{2k}$.

Next, the power spectra of scalar $P_\mathfrak{R}$ and tensor $P_g$ are defined as follows [20].

$$<\mathfrak{R}_k(\eta), \mathfrak{R}_l^*(\eta)> = \frac{2\pi^2}{k^3} P_\mathfrak{R} \delta^3(\bm{k} - \bm{l}), \tag{11}$$

$$<v_k(\eta) v_l^*(\eta)> = \frac{\pi m_P^2 a^2}{16 k^3} P_g \delta^3(\bm{k} - \bm{l}), \tag{12}$$

where $\mathfrak{R}_k(\eta)$ is the Fourier series of the curvature perturbation $\mathfrak{R}$ and $m_P$ is the Planck mass. The power spectra $P_\mathfrak{R}^{1/2}$ and $P_g^{1/2}$ are then written as follows [20].

$$P_{\mathfrak{R}}^{1/2} = \sqrt{\frac{k^3}{2\pi^2}} \left|\frac{u_k}{Z}\right|, \tag{13}$$

$$P_g^{1/2} = \frac{4\sqrt{k^3}}{m_P \sqrt{\pi} a} |v_k|. \tag{14}$$

Using the approximation of the Hankel function, the power spectra of the leading and next-leading corrections of $-k\eta$ in the case of the squeezed initial states (9) and (10) can be written as

$$P_{\mathfrak{R}}^{1/2} = (2^{-p}(-p)^p \frac{\Gamma(-p+1/2)}{\Gamma(3/2)} \frac{1}{m_P^2} \frac{H^2}{|H'|})|_{k=aH} (1-\frac{(-k\eta)^2}{2(1+2p)})|c_1 e^{-ip\pi/2} + c_2 e^{ip\pi/2}|$$

$$\cong (\frac{H^2}{2\pi\dot\phi})|_{k=aH} |c_1 e^{-ip\pi/2} + c_2 e^{ip\pi/2}|, \tag{15}$$

$$P_g^{1/2} = (\frac{2^{-p+1}}{\sqrt{\pi}}(-p)^p \frac{\Gamma(-p+1/2)}{\Gamma(3/2)} \frac{H}{m_P})|_{k=aH} (1-\frac{(-k\eta)^2}{2(1+2p)}) \times |c_{g1} e^{-ip\pi/2} + c_{g2} e^{ip\pi/2}|, \tag{16}$$

where $\Gamma(-p+1/2)$ represents the Gamma function and $P_g^{1/2}$ is multiplied by a factor of $\sqrt{2}$ for the two polarization states. These formulae differ slightly from Hwang's formula [21] due to the introduction of the term $e^{-ip\pi/2}$ into equation (8), as required such that in the limit $\eta \to -\infty$, the field $u_k(\eta)$ must approach plane waves. The quantities $C(k)$ and $C_g(k)$ are defined as

$$C(k) = c_1 e^{-ip\pi/2} + c_2 e^{ip\pi/2}, \tag{17}$$

$$C_g(k) = c_{g1} e^{-ip\pi/2} + c_{g2} e^{ip\pi/2}. \tag{18}$$

If $|C(k)| = 1$, the leading term of $-k\eta$ in equation (15) can be written as $P_{\Re}^{1/2} = (\frac{H^2}{2\pi\dot{\phi}})|_{k=aH}$ [24]. However, equations (17) and (18) imply that if, under some physical circumstances, the Bunch-Davies state is not adopted as the initial condition of the fields $u_k(\eta)$ or $v_k(\eta)$, and the possibility exists that $|C(k)| \neq 1$ or $|C_g(k)| \neq 1$. In section 3, the values of $|C(k)|$ and $|C_g(k)|$ are calculated using a number of pre-inflation models.

## 3. Calculation of power spectrum

The effect of the length of inflation is examined using simplified models of pre-inflation as an illustration. Here, the pre-inflation model is considered to consist simply of a radiation-dominated period or a scalar matter-dominated period. A simple cosmological model is assumed, as defined by

$$\text{Pre-inflation:} \quad a_P(\eta) = b_1(-\eta - \eta_j)^r,$$

$$\text{Inflation:} \quad a_I(\eta) = b_2(-\eta)^p, \tag{19a}$$

where

$$\eta_j = (\frac{r}{p} - 1)\eta_2, \quad b_1 = (\frac{p}{r})^r(-\eta_2)^{p-r}b_2. \tag{19b}$$

The scale factor $a_I(\eta)$ represents ordinary inflation. If $p < -1$, the inflation is power-law inflation ($p = -10/9$, $a(t) = t^{10}$), and if $p = -1$, the inflation is de-Sitter inflation, which is not considered here. Inflation is assumed to begin at $\eta = \eta_2$. In pre-inflation, for the case $r = 1$, the scale factor $a_p(\eta)$ indicates that pre-inflation is a radiation-dominated period, whereas for the case $r = 2$, the scale factor $a_p(\eta)$ indicates a scalar matter-dominated period.

Using the pre-inflation models and taking account of the matching conditions, the quantities $|C(k)|$ and $|C_g(k)|$ (the power spectra) are calculated as follows.

### 3.1 Radiation-dominated period before inflation

In the case of a radiation-dominated period before inflation, the scale factor becomes $a_p = b_1(-\eta - \eta_j)$, i.e., $r = 1$. A difference between the scalar and tensor perturbations (gravitational waves) occurs in the radiation-dominated period, that is, the equations for the fields become different. In the case of gravitational waves, the solution to equation (6) is written as $f_k^R(\eta) = \exp[-ik(\eta + \eta_i)] / \sqrt{2k}$. In the case of curvature perturbations, the field equation $u_k$ can be written as equation (6) with the value $c_s^2 = 1/3$ and $Z = a_p[2(\mathcal{H}^2 - \mathcal{H}')/3]^{1/2}/(c_s \mathcal{H})$ [23,26], where $\mathcal{H} = a_p'/a_p$. The solution is then written as $f_k^{SR}(\eta) = 3^{1/4} \exp[-ik(\eta + \eta_i)/\sqrt{3}] / \sqrt{2k}$. For simplicity, it is assumed that the mode functions of the radiation-dominated period can be written as $f_k^R(\eta)$ for gravitational waves, and as $f_k^{SR}(\eta)$ for scalar perturbations. In power-law inflation, the equations can be written as equation (7). The general mode functions in inflation can then be written as equations (9) and (10). The coefficients $c_1$, $c_2$, $c_{g1}$ and $c_{g2}$ are fixed using the matching condition in which the mode function and first $\eta$-derivative of the mode function are continuous at the transition time $\eta = \eta_2$ ($\eta_2$ is the beginning of inflation). The coefficients $c_1$, $c_2$, $c_{g1}$ and $c_{g2}$ can then be calculated analytically, and $C(k)$ and $C_g(k)$ can be derived from equations (17) and (18) as follows.

$$C(k) = -\frac{\sqrt{\pi}}{2 \cdot 3^{3/4}\sqrt{2z}} e^{iz/\sqrt{3}p}$$

$$\{(-3+3p+i\sqrt{3}\,z)(H^{(1)}_{-p+1/2}(z)+H^{(2)}_{-p+1/2}(z))+3z(H^{(1)}_{-p+3/2}(z)+H^{(2)}_{-p+3/2}(z))\}, \quad (20)$$

$$C_g(k) = -\frac{\sqrt{\pi}}{2\sqrt{2z}} e^{iz/p}$$

$$\{(-1+p+iz)(H^{(1)}_{-p+1/2}(z)+H^{(2)}_{-p+1/2}(z))+z(H^{(1)}_{-p+3/2}(z)+H^{(2)}_{-p+3/2}(z))\} \quad (21)$$

where $z = -k\eta_2$. As $z \to 0$, $|C(k)|^2$ and $|C_g(k)|^2$ become zero (i.e., $z^{-2p}$) in both cases, while as $z \to \infty$ (from large to small scales), the quantities $|C(k)|^2$ and $|C_g(k)|^2$ are approximately given by

$$|C(k)|^2 \cong \frac{1}{\sqrt{3}}(2+\cos(p\pi+2z)), \quad (22)$$

$$|C_g(k)|^2 \cong 1-\frac{p(p-1)\cos(p\pi+2z)}{2z^2}. \quad (23)$$

The behavior differs between the scalar and tensor cases. $|C(k)|^2$ oscillates around $2/\sqrt{3} \cong 1.1547$, but the amplitude does not depend on $p$ of the leading order. Numerically, $1/\sqrt{3} \leq |C(k)|^2 \leq \sqrt{3}$, but $|C_g(k)|^2$ becomes 1. The quantities $|C(k)|^2$ are plotted as a function of $z$ ($=-k\eta_2$) in figure 1 for the case $p = -500/499$. In the case of gravitational waves, the behavior of $|C_g(k)|^2$ is similar to that in the scalar matter-dominated case, which is shown in figure 2. The correct plot of $|C_g(k)|^2$ is given in Ref. [27].

These figures show plots for a length of inflation of 60 *e*-folds. Perturbations of the current Hubble horizon size are shown for multiples of this length of inflation, that is, if the length of inflation is *a* times longer, the perturbations of the current Hubble horizon size is given by $z \approx a$. In the case of the curvature perturbations, if a longer inflation exists, the quantity $|C(k)|^2$ behaves according to equation (22) from the super horizon scales to small scales, oscillating around $2/\sqrt{3}$ (see figure 1). It is important to note that even a very long inflationary period cannot remove the oscillation of $|C(k)|^2$ around $2/\sqrt{3}$.

The ratio of gravitational waves to the curvature perturbations on the power spectrum is given by

$$R(k) = \frac{P_g}{P_\mathfrak{R}} = \frac{16(p+1)}{p}\frac{|C_g(k)|^2}{|C(k)|^2}, \qquad (24)$$

where the term $16(p+1)/p$ represents the contribution of power-law inflation. In the case of $z \to 0$ (super large scales), $R_c(k) \approx 1/\sqrt{3}$ ($R_c = |C_g(k)|^2/|C(k)|^2$), and in the case of $|z| \gg 1$, $R_c(k)$ becomes $\sqrt{3}/(2 + \cos(p\pi + 2z))$.

## 3.2 Scalar matter-dominated period before inflation

When the period before inflation is dominated by scalar matter (given by the inflaton field ϕ), the scale factor becomes $a_p = b_1(-\eta - \eta_j)^2$, i.e., $r = 2$. In this case, equation (6) becomes the same for the scalar and tensor cases, and $C(k)$ and $C_g(k)$ take the same form. By a similar procedure as employed in section 3.1, $C(k)$ and $C_g(k)$ can be derived as follows.

$$C(k) = C_g(k) = \frac{-i\sqrt{\pi}}{8\sqrt{2z^3}} e^{2iz/p} \{(p^2 + 4z(i+z) - 2p(1+iz))(H^{(1)}_{-p+1/2}(z) + H^{(2)}_{-p+1/2}(z))$$

$$+ 2(p-2iz)z(H^{(1)}_{-p+3/2}(z) + H^{(2)}_{-p+3/2}(z))\}. \tag{25}$$

The quantities $|C(k)|^2$ are proportional to $z^{-2-2p}$ when $z \to 0$ ($k \to 0$), and so $|C(k)|$ becomes zero. For $z \to \infty$ (for large to small scales), $|C(k)|^2$ and $|C_g(k)|^2$ are obtained as

$$|C(k)|^2 = |C_g(k)|^2 \cong 1 - \frac{p(p-2)\cos(p\pi + 2z)}{4z^2}. \tag{26}$$

In this case, $|C(k)|^2 = |C_g(k)|^2 \cong 1$, which differs from the case for the radiation-dominated period before inflation.

To compare the case of the radiation-dominated period with the case of the scalar matter-dominated period, $|C(k)|^2$ ($=|C_g(k)|^2$) as given above is plotted as a function of $z$ in figure 2 for the case $p = -500/499$. The form undergoes very little change with $p$. As the quantities $C_g(k)$ and $C(k)$ are the same, the ratio $R_c = |C_g(k)|^2 / |C(k)|^2 = 1$, and the ratio $R(k) = 16(p+1)/p$.

### 3.3. Matching conditions

One of two matching conditions for scalar perturbations was used in sections 3.1 and 3.2 above. The first is a matching condition in which the gauge potential and its first η-derivative are continuous at the transition point (Condition A). This matching condition allows the initial condition of pre-inflation to be decided rationally, that is, in the limit $\eta \to -\infty$, the field $u_k(\eta)$ approaches plane waves. The second matching condition is that of Deruelle and Mukhanov [17] for cosmological perturbations, which requires that the transition occurs on a hyper-surface of constant energy (Condition B). $C(k)$ can be calculated under matching condition B by a similar method to that for matching condition A. This case of matching condition is investigated in detail in Ref. [27], but since the result is important, it is shown in the appendix. In this case, the two models (radiation- and scalar matter-dominated) exhibit similar behavior (figure 3), where $|C(k)|^2$ oscillates from large to small scales with large amplitude.

## 4. Angular power spectra

Calculation of the power spectra for scalar and tensor perturbations under the two models and matching conditions above for power-law inflation reveals two interesting properties. First, when the length of inflation is finite, the power spectra can be expressed by decreasing functions for $z \to 0$ (in the range of super large scales), and this behavior is seen in all of the cases considered. Second, under matching condition A in the case of a scalar matter-dominated pre-inflation period, the power spectrum becomes 1 as $z \to \infty$, and in the case of radiation-dominated pre-inflation period it oscillates from large to small scales, but the amplitude is nearly 1. Under matching condition B, the power spectra oscillate from large to small scales for both radiation- and scalar matter-dominated pre-inflation, and the amplitude is very large and becomes infinite as $p \to -1$ (de-Sitter inflation). These two properties can be observed from figures 1–3, and the details of these properties have been discussed in the preceding papers [14,15,27].

The effect of this behavior on the angular power spectra is investigated by comparison of the calculated angular power spectra with the WMAP data. The angular TT and TE power spectra $C_\ell$ and $C_\ell^{TE}$ can be written as [28]

$$C_\ell = 4\pi \int T_\Theta^2(k,\ell) \, P_{\mathfrak{R}}(k) \frac{dk}{k}, \tag{27}$$

$$C_\ell^{TE} = 4\pi \int T_\Theta(k,\ell) T_E(k,\ell) \, P_{\mathfrak{R}}(k) \frac{dk}{k}, \tag{28}$$

where $T_\Theta(k,\ell)$ and $T_E(k,\ell)$ are transfer functions. The angular TT and TE power spectra were computed using a modified CMBFAST code [18] assuming a flat universe and the following parameter values: baryon density $\Omega_b = 0.046$, dark energy density $\Omega_\Lambda = 0.73$, and present-day expansion rate $H_0 = 72$ km s$^{-1}$ Mpc$^{-1}$. The present treatment considers the length of inflation (*a*), the value of *p* $(a(\eta) = (-\eta)^p \approx t^q)$, and the reionization optical depth τ. The standard value of τ = 0.17 is used in many cases, and $a = 1$ indicates that inflation starts at the time when the perturbations of the current Hubble horizon size exceed the Hubble radius in inflation, that is, the length of inflation is assumed to be close to 60 *e*-folds. From equations (20), (21), (25), (A3) and (A6), the quantities $|C(k)|^2$ and $|C_g(k)|^2$ can be derived analytically using the Bessel function. In the present calculations of the angular power spectra, the expansions of the Bessel function in $|C(k)|^2$ and $|C_g(k)|^2$ at $z = 0$ and $z = \infty$ were used. The angular TT power spectra were normalized with respect to 11 data points in the WMAP data from $l = 65$ to $l = 210$ so as to average the small changes in $l$ due to the models and the contribution of oscillation. The same values are used in the analysis of the angular TE spectrum.

In the calculation of the angular power spectrum, the scalar-tensor ratio at $l = 2$ is required, which for power-law inflation is written as $R(k) = P_g / P_\Re = 16(p+1)/p |C_g(k)|^2 / |C(k)|^2$. For example, the value of $16(p+1)/p$ is 0.32 at $p = -50/49$. In the case of matching condition A, $R_c(k) = |C_g(k)|^2 / |C(k)|^2$ is close to 1 for radiation-dominated pre-inflation, and exactly 1 for scalar matter-dominated pre-inflation. However, in the case of matching condition B, $R_c(k)$ has a peculiar shape, approaching zero but displaying strong peaks at every $\pi$ value of $z$. For example, in the case of radiation-dominated pre-inflation, an approach to zero indicates $(1+p)/(2\sqrt{3}p)$ ($\cong 0.006$ at $p = -50/49$), and the height of the peak is $2\sqrt{3}p/(1+p)$ ($\cong 173$ at $p = -50/49$) (see figure 4). This effect is considered to be accounted for in the integration, since the average value of the integration from $z = 0$ to $z = 10$ is 0.64 at $p = -10/9$ and 0.055 at $p = -500/499$. Thus, in the combination of this effect, the scalar-tensor ratio is very small in the case of matching condition B. For example, at $p = -500/499$, $R(k) = (p+1)/p\, R_c(k) \approx 0.032 \times 0.06$. Therefore, in the case of matching B, the tensor part represents only a minor contribution to the angular spectrum.

In the likelihood analysis, the values of $\chi^2$ were calculated for the angular TT and TE power spectra against 899 and 449 WMAP data points using the WMAP Likelihood Code [19]. The angular TT and TE power spectra for the ΛCDM model were also calculated using the above parameters, with a spectral index of 1.0 and no running with de-Sitter-like inflation. The ΛCDM model with power-law inflation, i.e., the condition $|C(k)|^2 = |C_g(k)|^2 = 1$, was also considered with these parameter values. The magnitude of the angular TT power spectrum for power-law inflation at small $l$ ($l \leq 100$) changes with $p$, and is larger than that for the ΛCDM case at small values of $q$ ($a(\eta) = (-\eta)^p \approx t^q$). Therefore, to fit the WMAP data, the value of $p$ must be in the range $-300/299 \leq p < -1$, in which the total $\chi^2$ is smaller than for the ΛCDM model. The angular TE power spectrum retains almost the same form in all cases.

(1) *Radiation-dominated pre-inflation coupled with matching condition A*

The angular TT power spectrum for radiation-dominated pre-inflation coupled with matching condition A is shown in figure 5 for various values of $a$ at $p = -500/499$, with the ΛCDM model result shown for comparison. Three interesting results can be observed here. The first is that the length of inflation is critical. Suppression of the cosmic microwave background (CMB) power spectrum is obtained in the case of $a \approx 1$. At $a \leq 1.2$, the angular power spectrum of $l = 2,3$ fits the WMAP data within the limit of error. As the decrease occurs smoothly, it seems unlikely that only the value for $l = 2$ is very small, as shown by the WMAP data. Secondly, the power spectrum exhibits an oscillation from large to small scales (see figure. 1), and the influence of this oscillation on the angular power spectrum is very interesting. In figure 5, the angular TT power spectrum displays a small oscillation at $l \geq 5$, which may explain the small oscillation seen in the angular TT power spectrum of the WMAP data. Figure 5 also resolves a small change in the spectral shape, including the shape of the first peek, indicating a dependence on the value of $a$. Such behavior occurs at $a \approx 50$. The dependence on $p$ is illustrated in figure 6. To fit the WMAP data, $p$ must be in the range of $-300/299 \leq p \leq -1$, with a slight dependence on the value of $a$, indicating that the contribution of gravitational waves is very small. Within such a range of $p$, the shape of the spectrum remains largely unchanged with changes in $p$.

The angular TE power spectrum for $l \geq 20$ and $a \geq 10$ is similar to that for the $\Lambda$CDM model [19] in almost all cases considered here. However, a clear difference emerges for the finite inflation model ($a \approx 1$), where the value of $C_\ell^{TE}$ is smaller than in the $\Lambda$CDM model for the same value of $\tau$, and $C_\ell^{TE}$ oscillates with two peaks at $l = 4, 11$ (see figure 7). As the reionization optical depth changes from 0.17 to 0.25, $C_\ell^{TE}$ becomes larger (figure 7), accompanied by a slight improvement in $\chi^2$ for the TE spectrum but no appreciable change in total $\chi^2$.

By calculating $\chi^2$ for the TT and TE spectra over all cases, better models of radiation-dominated pre-inflation can be sought. If the TT power spectrum for $l = 2, 3$ is required to fit the WMAP data within the limit of error, that is, the case of $a \approx 1$, then better models are obtained for $a = 1.03 - 1.045$, as the difference in total $\chi^2$ is not large over this range ($\Delta\chi^2 \leq 2.86$). The best model has $a = 1.032$, giving $\chi^2(TT) = 982.97$, $\chi^2(TE) = 460.80$, and $\chi^2 = 1443.77$, which are slightly higher than the corresponding values obtained for the $\Lambda$CDM model ($\chi^2(TT) = 975.46$, $\chi^2(TE) = 455.91$, $\chi^2 = 1431.37$). Alternatively, if the angular TT power spectrum for $l = 2,3$ is not required to fit the WMAP data within the limit of error, there are some cases in which the value of $\chi^2$ is smaller than for the $\Lambda$CDM model, for example, when $a = 3.0$, $\chi^2(TT) = 977.16$, $\chi^2(TE) = 452.74$, $\chi^2 = 1429.90$ are obtained.

(2) *Scalar matter-dominated pre-inflation coupled with matching condition A*

The angular TT power spectrum in the case of scalar matter-dominated pre-inflation and matching condition A is shown in figure 8. Similar behavior to that seen here has been derived in many previous papers [5-10]. The length of inflation is again critical, and suppression of the angular power spectrum is obtained in the case of $a \approx 1$. The angular TT power spectrum for $l = 2,3$ fits the WMAP data within the limit of error at $a \leq 1.2$. As the decrease occurs smoothly, similar to the radiation-dominated case, it is unlikely that only the value for $l = 2$ is very small, as shown by the WMAP data. When the length of inflation is longer (see figure 8, $a = 3.0$), the spectrum becomes similar to that predicted by the $\Lambda$CDM model. The power spectrum in the matter-dominated case does not oscillate from large to small scales (see figure 2), and in all 4 scenarios ($a = 0.6, 1.0, 3.0$ and $\Lambda$CDM) the spectra become the same shape as that for $l \geq 100$ (see figure 8). This demonstrates that the small oscillation of the angular TT power spectrum in the radiation-dominated case is due to oscillation of the power spectrum. The dependence on $p$ is similar to that for the radiation-dominated case. To fit the WMAP data, $p$ must be in the range $-200/199 \leq p < -1$, with a slight dependence on the value of $a$, again indicating that the contribution of gravitational waves is very small. If $p$ is in the range $-500/199 \leq p < -1$ ($a = 0.8$), the total $\chi^2$ is smaller than for the $\Lambda$CDM model.

The derived angular TE power spectra for $l \geq 20$ and $a \geq 10$ are similar to those for the $\Lambda$CDM model in all cases. However, a clear difference again emerges for the finite inflation model ($a \approx 1$), where the value of $C_\ell^{TE}$ is smaller than that in the $\Lambda$CDM model for the same value of $\tau$, and $C_\ell^{TE}$ peaks at $l = 5$ (see figure. 9). As the reionization optical depth increases from 0.17 to 0.25, $C_\ell^{TE}$ becomes larger, accompanied by a slight improvement in $\chi^2$(TE) but no appreciable improvement in total $\chi^2$. This behavior differs from that in the $\Lambda$CDM model and the radiation-dominated pre-inflation model. More precise measurement of the TE spectrum is required in order to determine which is the better model.

In the likelihood analysis, if the angular TT power spectrum for $l = 2, 3$ is required to fit the WMAP data within the limit of error (i.e., the case of $a \approx 1$), the better models are those with $a = 0.8 - 1.2$. Total $\chi^2$ is better over this range than for the $\Lambda$CDM model. The best model has $a = 1.2$, which gives $\chi^2$(TT) = 973.16, $\chi^2$(TE) = 455.78, and $\chi^2$ = 1428.94.

(3) *Matching condition B*

Under matching condition B, similar behavior of the angular TT power spectrum is obtained for both the radiation-dominated and scalar matter-dominated models. Figure 10 shows the results for $a$ = 1.0, 10.0 and 25.0 at $p = -500/499$ and for the ΛCDM model for the scalar matter-dominated case. The suppression of the angular TT power spectrum at $a \approx 1$ does not occur. Although the power spectra oscillate appreciably from large to small scales under this matching condition (see figure 3), the oscillatory behavior in figure 10 is not so prominent. However, as a large hump appears in the range $5 \leq l \leq 20$ at $a \approx 1$, the angular TT power spectrum in this case does not fit the WMAP data. At $a \geq 10$, the angular TT power spectra do not decrease at $l$ = 2,3, and oscillatory behavior occurs around the spectrum for the ΛCDM model. Although $\chi^2$ in many cases is considerably worse than under matching condition A, the model with $a$ = 25.0 yields $\chi^2(TT) = 973.63$, $\chi^2(TE) = 457.49$, and $\chi^2 = 1431.12$. TT and total values are thus slightly better than for the ΛCDM model. The TE power spectrum does not differ appreciably from that for the ΛCDM model, but similar to the case of matching condition A at $l \leq 20$, a slight oscillation is observed around the spectrum of the ΛCDM model. A similar *p*-dependence is also derived.

The angular TT and TE power spectra given by the best models identified above scalar matter- and radiation-dominated cases are plotted in figure 11 for comparison with those for the ΛCDM model.

## 5. Discussion and summary

The effect of the length of inflation was investigated by analyzing two pre-inflation models in which pre-inflation is either radiation-dominated or scalar matter-dominated. The effect of pre-inflation was described by the factor $|C(k)|$ (or $|C_g(k)|$), allowing the familiar formulation of the derived power spectrum of curvature perturbations or gravitational waves to be simply multiplied by this factor. These factors were calculated considering two matching conditions, and the characteristics of each of the scenarios were derived [13,14]. The angular TT and TE power spectra were calculated for various lengths of inflation, values of $p$, and reionization optical depths, and a likelihood analysis was performed to evaluate the fit to the WMAP data and facilitate comparison with the ΛCDM model.

The analysis reveals two important problems; the suppression of the TT spectrum at $l = 2,3$, and the oscillatory behavior of the TT spectrum. In the previous papers [13,14], the power spectrum was found to decrease on super large scales in all cases. Here, under matching condition A (the gauge potential and its first $\eta$-derivative are continuous at the transition point), the angular TT power spectra decreases at $l = 2,3$ near $a \approx 1$, and the angular TT power spectra fit the WMAP data within the limit of error at $a \leq 1.2$ (see figures. 5, 6 and 8). This type of behavior has been found in many studies in which power spectra similar to the scalar matter-dominated case are assumed or derived [5-10]. However, it appears difficult to obtain a decrease in the TT angular power spectrum under Deruelle and Mukhanov's matching condition (matching condition B, see figure 10).

Oscillation from large to small scales occurs in the power spectrum of curvature perturbations in the radiation-dominated case under matching condition A, and this oscillation was shown to induce a weak oscillation in the angular TT power spectrum accompanied by a slight change in spectral shape (including the first peak) depending on the length of inflation (see figure 5). Even at large values of $a$ (e.g., $a \approx 50$), these properties are retained. If the WMAP data exhibit oscillation, this pre-inflation model may be appropriate. Under matching condition B, the oscillation of $|C(k)|$ is very large (see figure 3), yet does not give rise to strong oscillation of the angular TT power spectrum (see figure 10). In comparison with matching condition A, the shape of the angular TT power spectrum exhibits a strong dependence on $a$.

To fit the WMAP data, both the pre-inflation models require $p$ to be in the range $-300/199 \leq p < -1$ ($q \geq 300$) under both matching condition A (see figure 6) and matching condition B. The $\Lambda$CDM model with power-law inflation yields similar behavior. This restriction on the value of $p$ is due to the contribution of gravitational waves, which must be very small, suggesting de-Sitter-like inflation. Therefore, $p$ values of $-300/299 \leq p < -1$ are required in order to fit the WAMP data. A similar result was derived in Ref. [28], although the obtained range of $p$ was slightly different (i.e., $-1.019 < p < -1$).

The angular TE power spectrum exhibits similar behavior to that of the $\Lambda$CDM model under matching condition A with $l \geq 20$ and $a \geq 10$, while a small difference is seen under matching condition B. However, with $l \leq 20$ and $a \approx 1$, distinct differences in behavior occur for both the radiation- and scalar matter-dominated cases (see figures 7 and 9). In the present models, the value of the TE power spectrum is smaller than for the ΛCDM model with $l \leq 20$ and the same value of $\tau$. The origin of this discrepancy can be seen in equations (27) and (28), which show that the angular TT and TE spectra include the same contribution of $P_{\mathfrak{R}}$: if the power spectrum $P_{\mathfrak{R}}$ decreases at super and large scales, then both TT and TE spectra decrease. Therefore, if only the value of the TE spectrum is to be increased, then the value of $\tau$ must be increased [1,7].

In summary, the effects of the length of inflation, the *p*-dependence of inflation, and $\tau$ on the angular TT and TE power spectra were examined for two pre-inflation models and two matching conditions. While the suppression of the spectrum at $l = 2,3$ can be explained by a model with finite length of inflation (near 60 *e*-folds with $q \geq 300$) and a scalar matter-dominated pre-inflation period, as has been derived in many previous papers [5-10], the present study newly shows that the same result can be explained by a model with a radiation-dominated pre-inflation period under otherwise identical conditions, and that the TT power spectrum exhibits oscillatory behavior. Although the three better models (the scalar matter- and radiation-dominated pre-inflation models and the ΛCDM model) do not differ substantially in figure 11, the value of $\chi^2$ for the radiation-dominated pre-inflation models remains relatively poor. It thus appears that a model with radiation-dominated pre-inflation does not satisfactorily explain the WMAP data in the parameter region considered in this study. As there are few simple models that express both spectral suppression at $l = 2,3$ and oscillatory behavior of the TT power spectrum, and the shape of the angular TT power spectrum changes subtly according to small variations in *a*, it remains desirable to identify parameter regions in which both properties and good $\chi^2$ values are achieved. For example, in the case of matching condition B, many cases do not fit the WMAP data, but one good case is obtained at $a = 25.0$. It was shown that the angular TE power spectrum exhibits some behaviors that differ from the $\Lambda$CDM model and the present pre-inflation models at small values of $l$. Determination of which of these three models provides the best fit for the WMAP data will thus require more accurate measurement of the TE spectrum at $l \leq 20$. From study of the $p$-dependence, de-Sitter-like inflation may be preferable.

Although the present models do not have a concrete physical basis, the method employed here appears to be applicable to any physical inflation model with finite length, and similar results are likely to be derived. As an example of a concrete physical model, our group is currently calculating the angular power spectrum for an inflation model described in terms of supergravity, incorporating target-space duality and nonperturbative gaugino condensation in the hidden sector [30], although this example differs from the scenarios considered in the present paper (i.e., the effect of pre-inflation is not considered). A similar analysis is also being investigated for the slow-roll inflation model with finite length.

**Acknowledgments**

The authors would like to thank the staff of Osaka Electro-Communication University for valuable discussions.

**Appendix**

The quantity $|C(k)|$ is calculated here for Deruelle's matching condition, which dictates that $\Phi$ and $\xi$ ($\mathfrak{R}$) are continuous at the transition time $\eta_2$, and the difference between the two matching conditions is investigated. The parameters $\Phi$ and $\xi$ can be written as [21,26]

$$\Phi = (-\mathcal{H}^2 + \mathcal{H}')\, (\frac{u}{Z})' / (c_s^2 k^2 \mathcal{H}), \tag{A1}$$

$$\zeta = (\mathcal{H}\Phi' - \Phi \mathcal{H}' + 2\mathcal{H}^2 \Phi)/(\mathcal{H}^2 - \mathcal{H}'), \tag{A2}$$

where $Z$ is written as $a(\eta)(\mathcal{H}^2 - \mathcal{H}')^{1/2}/(c_s \sqrt{4\pi G}\, \mathcal{H})$. As $\Phi$ and $\xi$ can be written in terms of $u_k(\eta)$, the coefficients $c_1$ and $c_2$ are obtained as follows. In the period of pre-inflation, the mode function $u_k(\eta)$ is derived from equation (6), and $\Phi_P$ and $\xi_P$ can be obtained using the relations (A1) and (A2). On the other hand, in inflation, $u_k(\eta)$ is expressed as $c_1 f_k^I(\eta) + c_2 f_k^{I*}(\eta)$, which is used to calculate $\Phi_I$ and $\xi_I$. From the relations $\Phi_P(\eta_2) = \Phi_I(\eta_2)$ and $\xi_P(\eta_2) = \xi_I(\eta_2)$, the coefficients $c_1$ and $c_2$ can be fixed. Here, the matching condition in which $\Phi$ and $\xi$ ($\mathfrak{R}$) are continuous at the transition point is adopted. However, the matching condition of Deruelle and Mukhanov must be written such that $\Phi$ and $\xi + k^2 \Phi/(3(\mathcal{H}^2 - \mathcal{H}'))$ are continuous at the transition point. In the present case of ($k\eta_2 = -z$), the value of $\xi$ becomes smaller than that of the $k^2$ term, causing the latter to dominate at $|z| \gg 1$. However, the coefficients $c_1$ and $c_2$ cannot be fixed using $\Phi$ and the $k^2$ term. Thus, the matching condition in which $\Phi$ and $\xi$ ($\mathfrak{R}$) are continuous at the transition point is adopted for the present treatment.

For the case of the scalar matter-dominated period before inflation, the coefficients $c_1$ and $c_2$ can be calculated analytically and $C(k)$ can be written from equation (17) as

$$C(k) = \frac{\sqrt{\pi}}{16\sqrt{3p(p+1)}\sqrt{z^3}}$$

$$i e^{2iz/p} \{(p^3 + p^2(-4-2iz) - 8iz + p(4+8iz-12z^2))(H^{(1)}_{-p+1/2}(z) + H^{(2)}_{-p+1/2}(z))$$

$$- 4(1+p)(p-2iz)z(H^{(1)}_{-p+3/2}(z) + H^{(2)}_{-p+3/2}(z))\}. \tag{A3}$$

In the case of $z \to 0$,

$$|C(k)|^2 \cong \frac{2^{-7+2p}\pi p(-2+p)^4 z^{-2-2p}}{3(1+p)(\Gamma(\frac{3}{2}-p))^2} + \frac{2^{-7+2p}\pi(-2+p)^3(24-28p-34p^2+p^3)\,z^{-2p}}{3p(1+p)(-3+2p)(\Gamma(\frac{3}{2}-p))^2}.$$

$$\tag{A4}$$

In the case of $z \to \infty$,

$$|C(k)|^2 \cong \frac{(4+8p+13p^2) + (4+8p-5p^2)\cos\theta}{12p(1+p)} + \frac{(-4-9p^2+5p^3)\sin\theta}{12(1+p)z}$$

$$+ \frac{(-2+p)(-4-16p+9p^2+(-4-12p-15p^2+10p^4)\cos\theta)}{48(1+p)z^2} \tag{A5}$$

where $\theta = p\pi + 2z$. $|C(k)|^2$ is plotted as a function of $z$ in figure 3 for $p = -500/499$. $|C(k)|^2$ becomes zero in the $z \to 0$ limit, and oscillates around $2(p+1)/3p \leq |C(k)|^2 \leq 3p/(2(p+1))$ in the $z \to \infty$ limit, the latter corresponding to a numerical range of $0.0013 \leq$ $|C(k)|^2 \leq 750$ for $p = -500/499$. $|C(k)|^2$ is large and exhibits oscillation in figure 3, yet maintains a constant value of 1 in figure 2.

For the case of the radiation-dominated period before inflation, the function $C(k)$ differs slightly from that in the scalar-matter-dominated case, but the behavior remains similar. Therefore, it is only necessary to obtain $C(k)$, as follows.

$$C(k) = \frac{\sqrt{\pi}}{43^{3/4}\sqrt{p(p+1)}\sqrt{z}} e^{iz/\sqrt{3}p} \{(\sqrt{3}+4\sqrt{3}\ p^2 - p(\sqrt{3}+6iz))$$
$$\times (H^{(1)}_{-p+1/2}(z) + H^{(2)}_{-p+1/2}(z)) - \sqrt{3}\ (1+p)z(H^{(1)}_{-p+3/2}(z) + H^{(2)}_{-p+3/2}(z))\}. \tag{A6}$$

For further details please see Ref. [27].

The behavior of $|C(k)|^2$ differs considerably between the two matching conditions, and the effect of the factor $(p + 1)$ in the denominator appears to be the main cause. It should also be noted that matching condition B presents a problem in that the case $p = -1$ (de-Sitter inflation) is not applicable for this condition. In de-Sitter inflation, the value of $\Phi$ is exactly zero, and in pre-inflation is non-zero. Therefore, a matching value of $\Phi$ between pre-inflation and de-Sitter inflation cannot be found. This fact also demonstrates the effect of the factor $p + 1$ of the denominators in equations (A3) and (A5). This problem only occurs in the case of matching $\Phi$ and $\xi$, and does not occur in the case of matching the gauge potential $u_k(\eta)$.

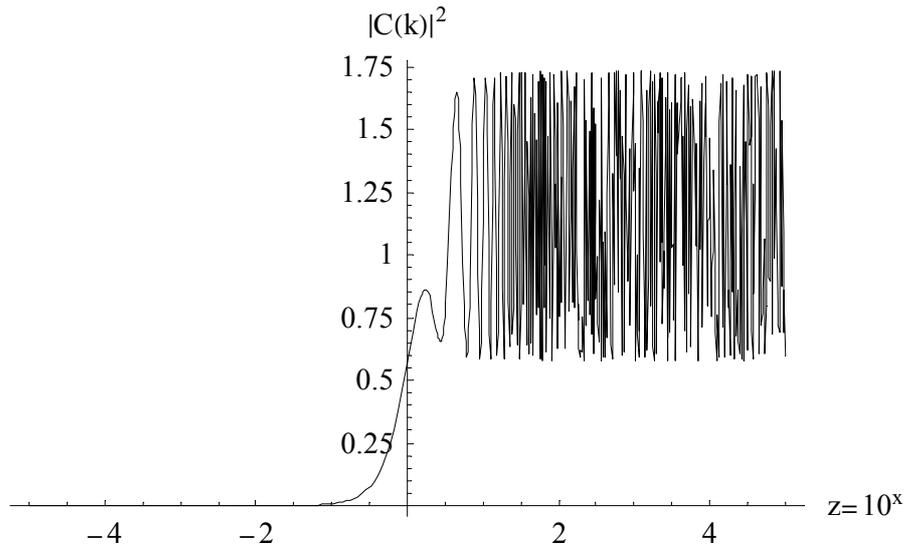

Figure 1. Factor $|C(k)|^2$ as a function of $z$ $(=-k\eta_2)$ for $10^{-5} \leq z \leq 10^5$ in the case of a radiation-dominated period before inflation under matching condition A ($p = -500/499$)

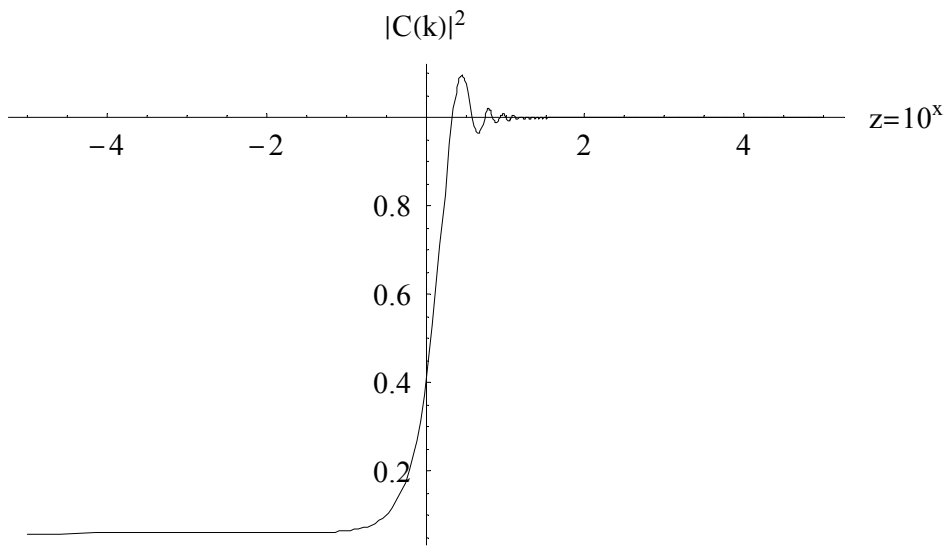

Figure 2. Factor $|C(k)|^2 = |C_g(k)|^2$ as a function of $z$ $(=-k\eta_2)$ for $10^{-5} \leq z \leq 10^5$ in the case of a scalar matter-dominated period before inflation under matching condition A ($p = -500/499$)

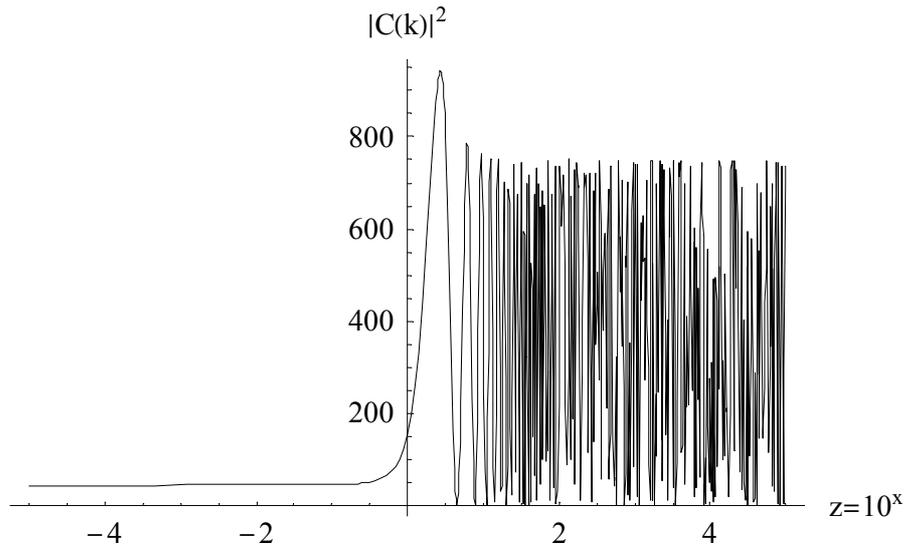

Figure 3. Factor $|C(k)|^2$ as a function of $z\ (=-k\eta_2)$ for $10^{-5} \leq z \leq 10^5$ in the case of a scalar matter-dominated period before inflation under matching condition B ($p = -500/499$)

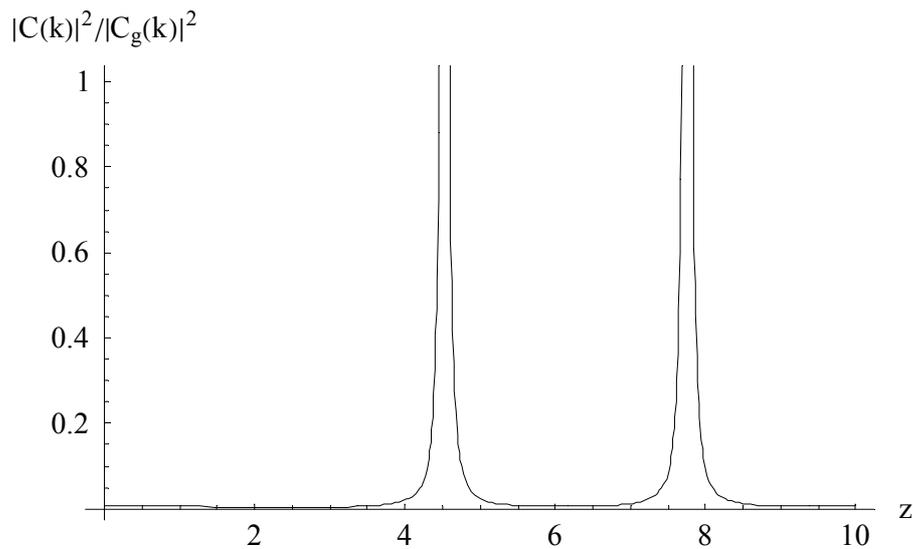

Figure 4. Factor $|C_g(k)|^2/|C(k)|^2$ as a function of $z\ (=-k\eta_2)$ for $0 \leq z \leq 10$ in the case of a radiation-dominated period before inflation under matching condition B ($p = -50/49$)

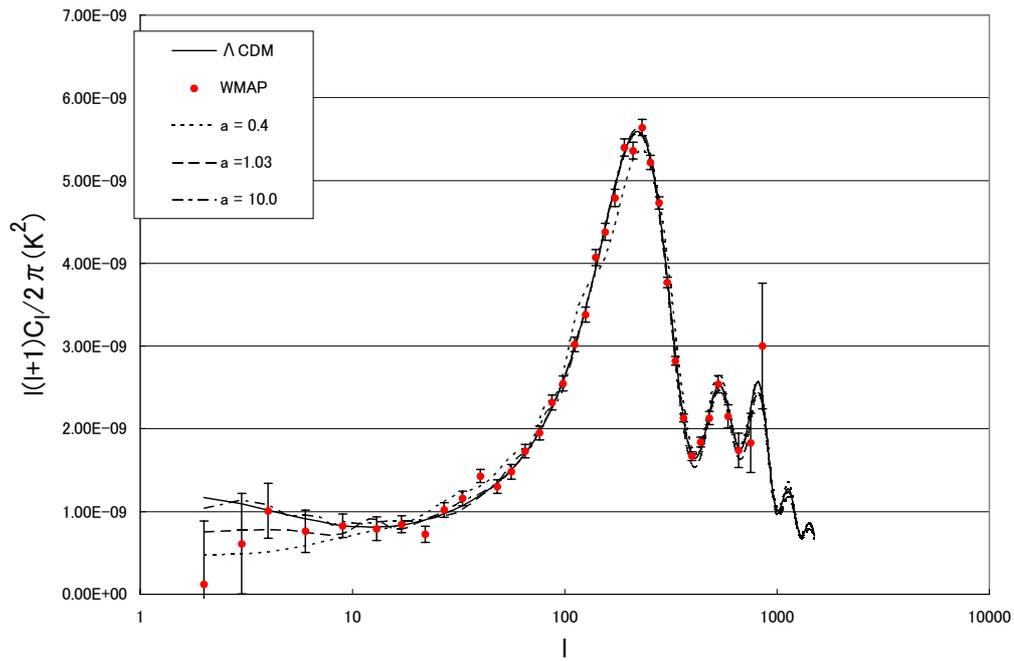

Figure 5. Angular TT power spectrum in the case of a radiation-dominated period before inflation under matching condition A for $a$ = 0.4, 1.03 and 10.0 ($p = -500/499$) with the ΛCDM model shown for comparison

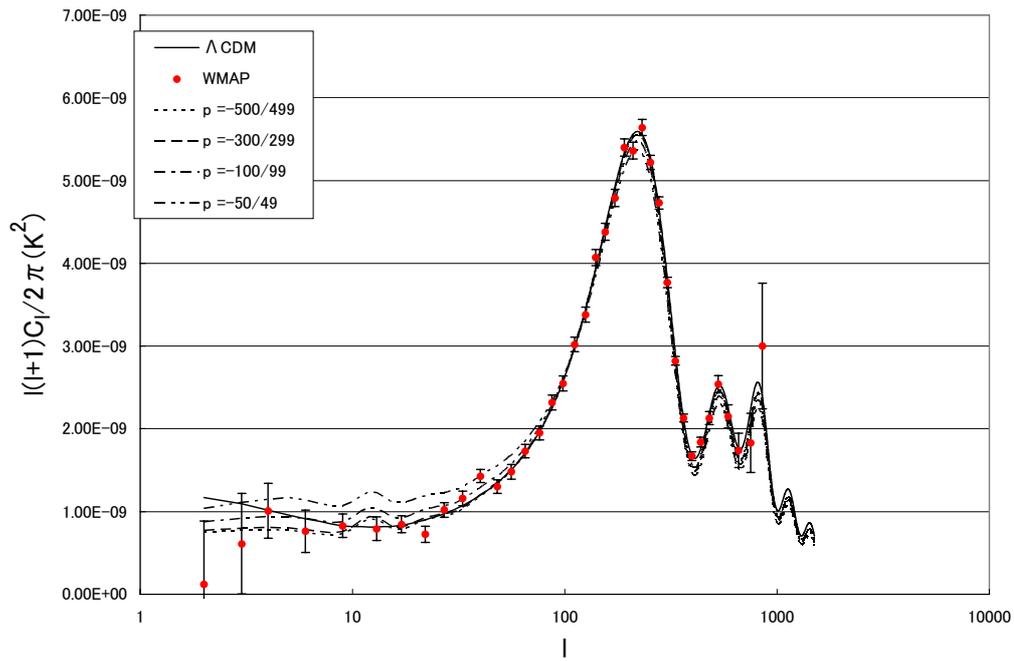

Figure 6. Angular TT power spectrum in the case of a radiation-dominated period before inflation under matching condition A for $p = -500/499, -300/299, -100/99$ and $-50/49$ ($a = 1.0$) with the $\Lambda$CDM model shown for comparison

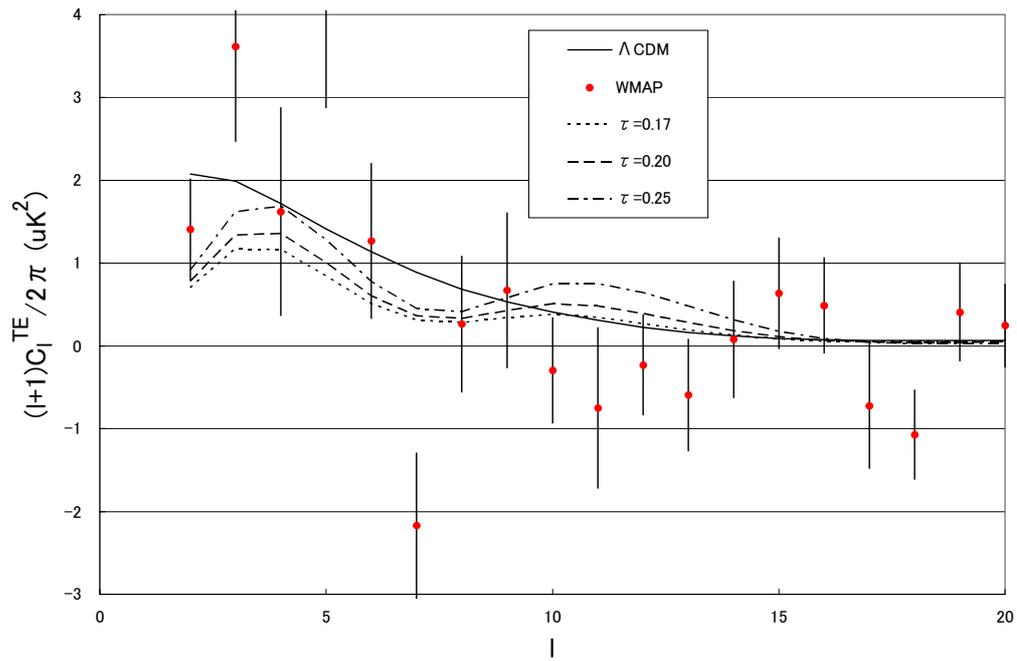

Figure 7. Angular TE power spectrum in the case of a radiation-dominated period before inflation under matching condition A for $\tau = 0.17, 0.20$ and $0.25$ ($p = -500/499$) with the $\Lambda$CDM model shown for comparison

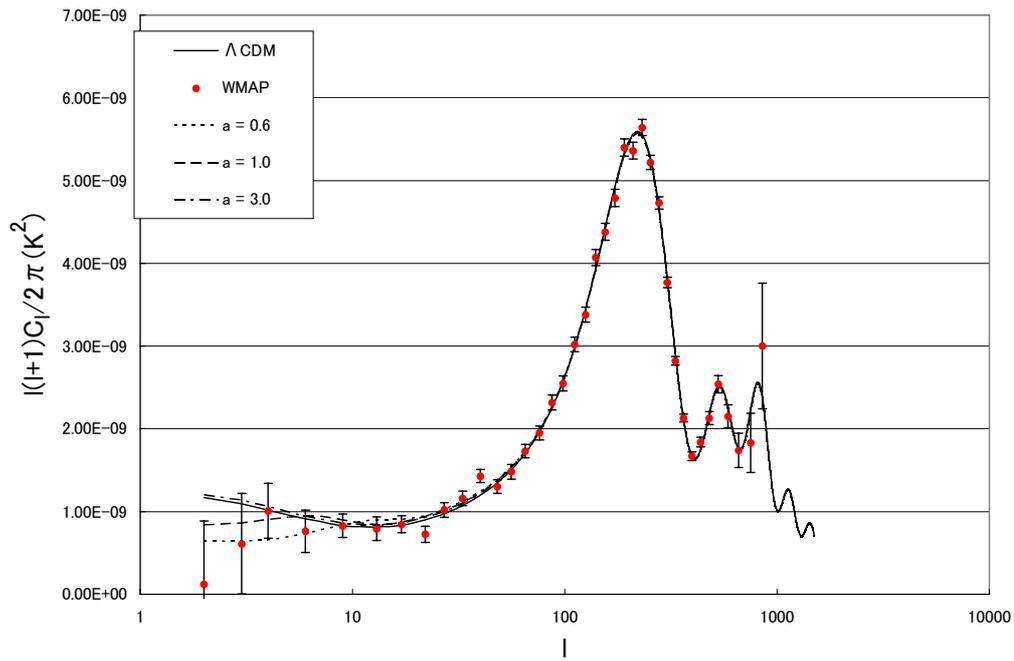

Figure 8. Angular TT power spectrum in the case of a scalar matter-dominated period before inflation under matching condition A for $a$ = 0.6, 1.0 and 3.0 ($p = -500/499$) with the ΛCDM model shown for comparison

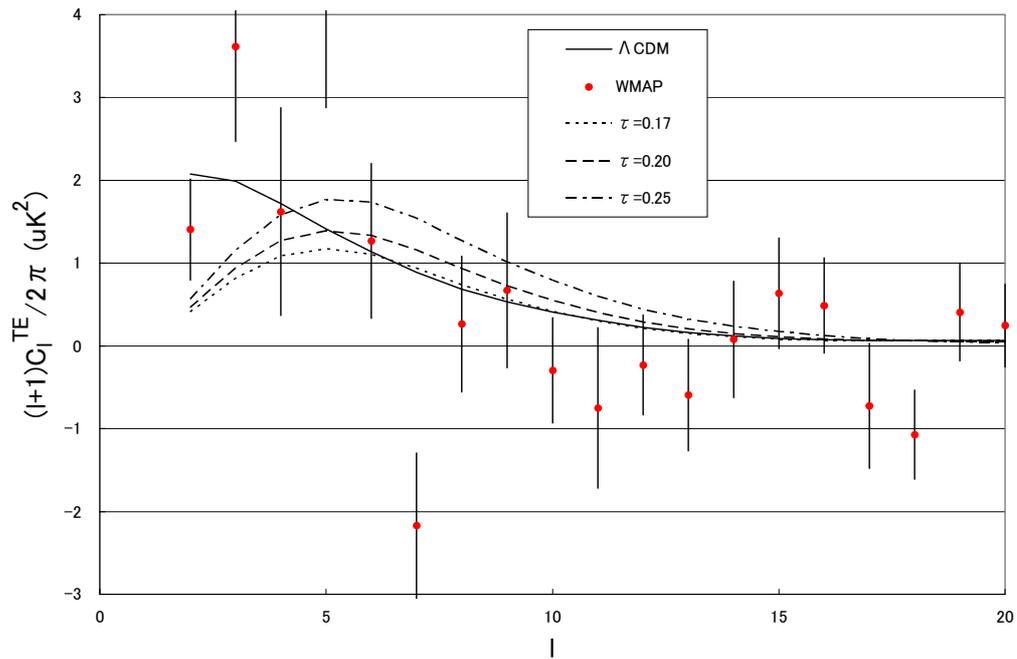

Figure 9. Angular TE power spectrum in the case of a scalar matter-dominated period before inflation under matching condition A for $\tau = 0.17$, $0.20$ and $0.25$ ($p = -500/499$) with the $\Lambda$CDM model shown for comparison

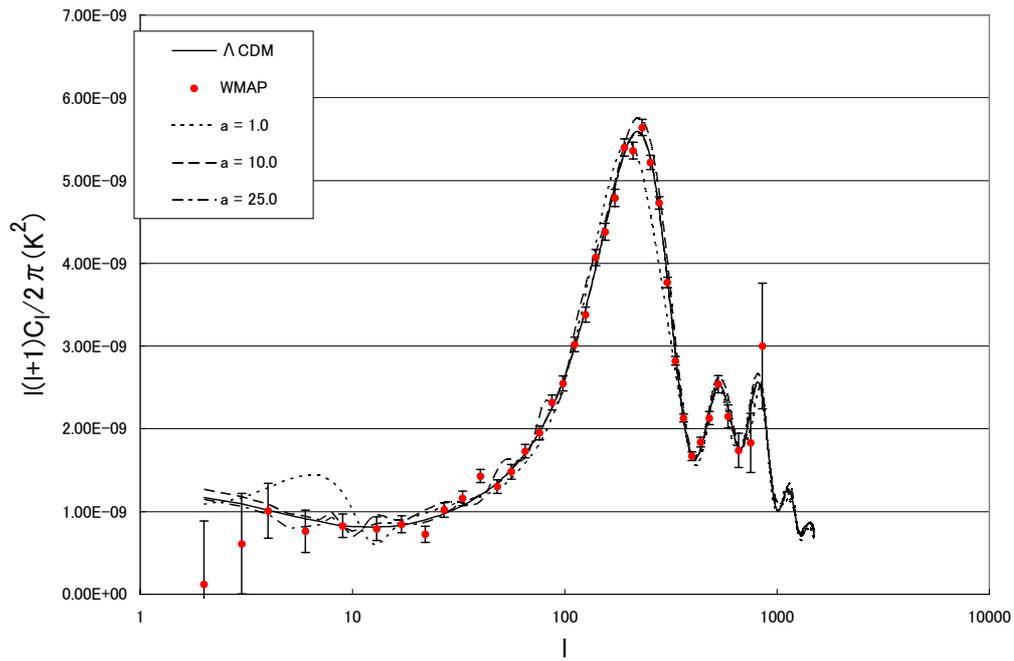

Figure 10. Angular TT power spectrum in the case of a scalar matter-dominated period before inflation under matching condition *B* for *a* = 1.0, 10.0 and 25.0 ($p = -500/499$) with the ΛCDM model shown for comparison

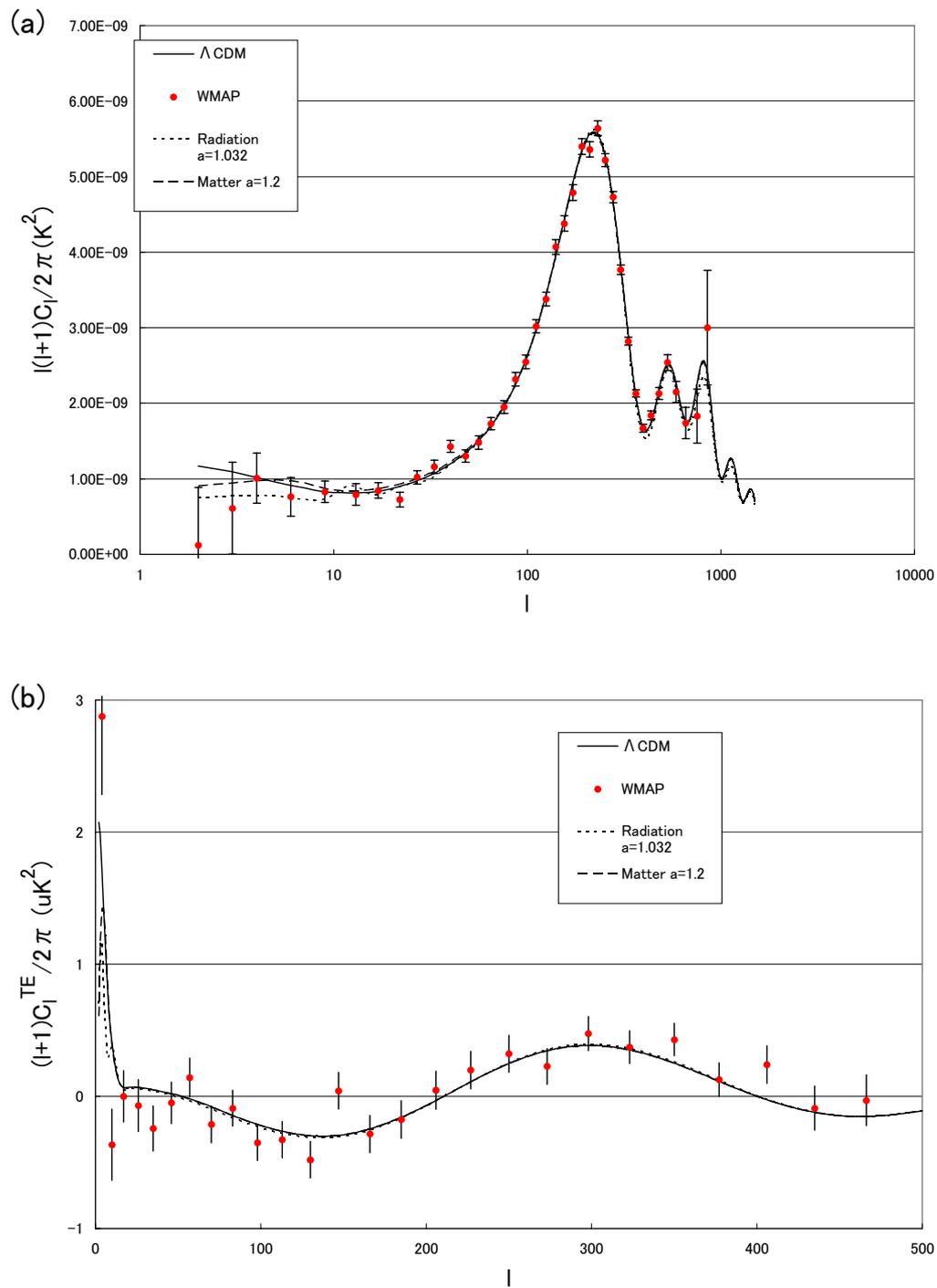

Figure 11. (a) Angular TT and (b) TE power spectra of the two best models for scalar matter-

($a = 1.2$) and radiation- ($a = 1.032$) dominated periods before inflation under matching condition A ($p = -500/499$) with the ΛCDM model shown for comparison